\documentclass[%
reprint,
amsmath,amssymb,
aps,
prl,
]{revtex4-2}

\usepackage[
  paperwidth=210mm,   
  paperheight=297mm,  
  top=20mm,           
  bottom=20mm,        
  left=17mm,          
  right=15mm,         
  columnsep=5mm      
]{geometry}

\usepackage{graphicx}
\usepackage{dcolumn}
\usepackage{bm}
\usepackage{xcolor}
\usepackage{booktabs} 
\usepackage{placeins}

\usepackage{comment}
\usepackage[font={scriptsize}, justification={justified},singlelinecheck=false]{caption}

\usepackage{ragged2e}

\makeatletter
\long\def\@makecaption#1#2{%
  \par
  \begingroup
    \footnotesize
    \setlength{\parindent}{0pt}
    \justifying
    \noindent #1.~#2\par
  \endgroup
}
\makeatother

\usepackage{graphicx}
\usepackage{amsmath}

\renewcommand{\selectlanguage}[1]{}
\usepackage{hyperref}
\hypersetup{
    colorlinks=true,
    linkcolor=blue,
    citecolor=blue,
    urlcolor=blue
}

\begin{document}

\title{\Large A thorium-229 optical nuclear clock with feedback loop}

\author{L. Toscani De Col$^1$} 
\thanks{These authors contributed equally to this work.}
\author{T. Riebner$^{1,2}$}
\thanks{These authors contributed equally to this work.}
\author{I. Morawetz$^1$}
\author{F. Schneider$^1$}
\author{N. Sempelmann$^1$}
\author{J. Schlachet-L\'epinay$^1$}
\author{F. Schaden$^1$}
\author{M. Bartokos$^1$}
\author{G. A. Kazakov$^{1,3}$}
\author{K. Beeks$^1$}
\author{B. Gerstenecker$^1$}
\author{M. Pimon$^1$}
\author{S. Lahs$^1$}

\author{A. Hellerschmied$^2$}
\author{T. Lercher$^2$}
\author{J. Premper$^2$}
\author{A. Niessner$^2$}
\author{M. Matus$^2$}

\author{H. Denker$^4$}

\author{M. Cizek$^5$}
\author{O. Cip$^5$}

\author{V. Lal$^6$}
\author{G. Zitzer$^6$}
\author{V. Petrov$^7$}
\author{J. Tiedau$^6$}
\author{M. V. Okhapkin$^6$}
\author{E. Peik$^{6}$}
\email{ekkehard.peik@ptb.de}
\author{T. Schumm$^{1}$} 
\email{thorsten.schumm@tuwien.ac.at}
\affiliation{\vspace{0.1cm}}
\affiliation{$^1$Vienna Center for Quantum Science and Technology, Atominstitut, TU Wien, Vienna, Austria}
\affiliation{$^2$Bundesamt für Eich- und Vermessungswesen (BEV), Vienna, Austria}
\affiliation{$^3$Wolfgang Pauli Institut, Vienna, Austria}
\affiliation{$^4$Institut für Erdmessung, Leibniz Universit\"at Hannover, Germany}
\affiliation{$^5$Institute of Scientific Instruments of the CAS, v.v.i., Brno, Czech Republic}
\affiliation{$^6$Physikalisch-Technische Bundesanstalt (PTB), Braunschweig, Germany}
\affiliation{$^7$Max-Born-Institute for Nonlinear Optics and Ultrafast Spectroscopy, Berlin, Germany}

\date{\today}

\begin{abstract}

The laser-accessible nuclear transition in the thorium-229 isotope has been identified as a promising candidate for the realization of an optical nuclear clock~\cite{Peik:2003}. Such a nuclear clock might rival or outperform current optical clocks based on electron-shell transitions in atoms or ions~\cite{Ludlow2015}, is expected to be more robust against external perturbations~\cite{Rellergert2010, Kazakov:2012}, and provides enhanced sensitivity in clock-based tests of fundamental principles of physics~\cite{Flambaum2006,Peik:2021}. Here, we implement a thorium-229 nuclear clock by stabilizing a continuous-wave laser to the 148\,nm nuclear transition with rapid feedback based on continuous absorption spectroscopy~\cite{morawetz2026continuous}. The thorium-229 nuclei are embedded into a millimeter-sized, room temperature calcium fluoride crystal. A subharmonic of the 148\,nm radiation is continuously compared to a Yb$^+$ single-ion clock. The nuclear clock shows a simple shot-noise limited scaling of the fractional frequency instability of $3\cdot10^{-12}/\sqrt{\tau/\text{s}}$ where $\tau$ is the averaging time, approaching $10^{-15}$ instabilities over 1\,day of continuous operation. Improvements of the instability by several orders of magnitude can be projected for future solid-state nuclear clocks. We use the nuclear clock to constrain models of ultralight dark matter by searching for periodic fluctuations and slow drifts in the nuclear transition energy, on time scales between 20\,s and 1\,day. Drawing benefit from the enhanced sensitivity of the thorium-229 transition, these constraints compete with the best atomic clocks concerning dark matter coupling to photons and go beyond previous measurements regarding coupling to the strong force and quarks.

\end{abstract}

\maketitle

Over the last 70 years, atomic clocks based on microwave and optical electronic transitions have successfully served as frequency standards based on their high levels of stability and accuracy \cite{Vanier_2005,Wynands_2005,Ludlow2015,Fortier:26}. To elevate these atomic timekeeping devices to the next generation, in 2003, Peik and Tamm~\cite{Peik:2003} proposed a nuclear clock based on the extraordinarily low-energy isomeric state of the thorium-229 (Th-229) isotope at 8.4\,eV. With a lifetime approaching 1\,h for a bare nucleus in vacuum, this transition offers a potential resonance quality factor on the order of $10^{19}$~\cite{Campbell:2012}. Additionally, the nucleus couples only weakly to perturbative fields~\cite{Beeks:2021}, which allows the construction of an optical clock implemented in a solid material at room temperature~\cite{Zhang:2024}. By doping microgram amounts of Th-229 into a crystal host, where the nuclei are confined to the Lamb-Dicke regime, an experimentally simple and robust clock can be created, offering a large gain in signal-to-noise ratio compared to approaches based on single or few isolated nuclei, while moderately degrading the resonance quality factor~\cite{Peik:2003,Campbell:2012,Kazakov:2012,ooi2026frequency}. 
Apart from these technical advantages, a clock based on the Th-229 nuclear transition is of great interest for testing theories beyond the standard model of particle physics that suggest the time variation of fundamental constants~\cite{Flambaum2006}. The low energy of the Th-229 isomer transition originates from a coincidental near-cancellation of the MeV-scale Coulomb and nuclear contributions to the binding energies of the Th-229 ground and excited state, making the transition frequency highly sensitive to fluctuations in the coupling constants of the involved interactions~\cite{Flambaum2006,Peik:2021,Fine_structure_Beeks_2025}. As a result, already a solid-state nuclear clock in its development phase will be competitive with or may even surpass the sensitivity of advanced atomic clock comparisons in the search for dark matter.
The realization of a nuclear clock builds upon half a century of scientific discoveries. Initiated by Kroger and Reich in 1976~\cite{Kroger76}, the energy of the Th-229 nuclear transition was narrowed down from an originally conjectured value below 100\,eV, via $-$1(4)\,eV~\cite{burke1990additional} and 3.5(1.0)\,eV~\cite{Helmer1994} to 7.8(5)\,eV~\cite{Beck2007, Beck09} using gamma spectroscopy methods. Using the internal conversion decay channel, the existence of the Th-229 isomer was proven in 2016~\cite{VonDerWense2016} and the energy value refined to 8.28(17)\,eV~\cite{Seiferle:2019}. Ultimately, an optical detection of the Th-229 radiative nuclear decay was realized in 2022, determining the isomer energy to 8.338(24)\,eV~\cite{Kraemer:2022} in the vacuum-ultraviolet (VUV) range. Thereafter, resonant laser excitation in Th:CaF$_2$, Th:LiSrAlF$_6$ crystals, and in ThF$_4$ thin films was reported in 2024~\cite{Tiedau:2024, Elwell:2024, Zhang:2024b}, quickly followed by the resolution of the nuclear quadrupole structure in Th:CaF$_2$~\cite{Zhang:2024,hiraki2025laser}. 
\begin{figure*}[t!]
\centering
\includegraphics[width=1\textwidth]{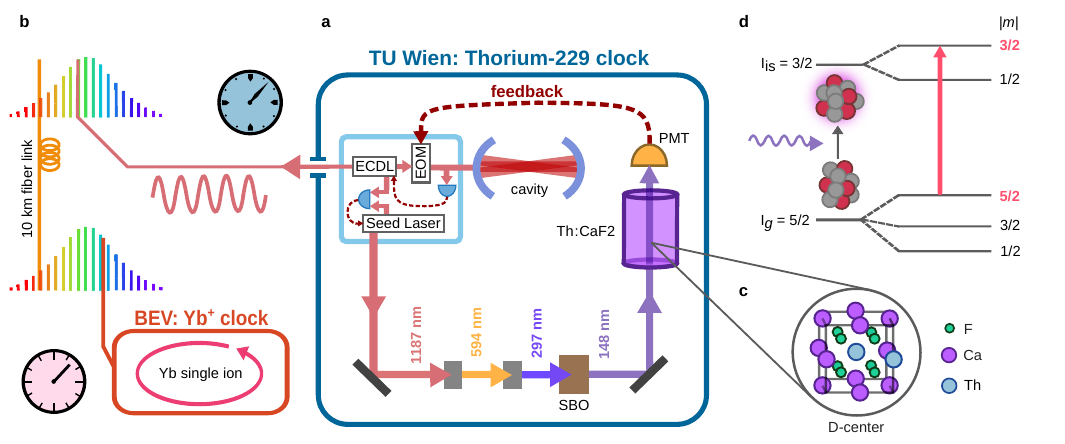}
\caption{\textbf{Schematic representation of the setup. a}, Thorium-229 nuclear clock located at TU Wien. It is comprised of an ECDL, which serves as a clock laser, stabilized to a high-finesse cavity via the sideband of an EOM, and a seed laser which is frequency-quadrupled, followed by the last SBO-based VUV frequency generation doubling. The seed laser is modulated via the offset lock to generate the error signal from the absorption of the thorium-229 nuclei in the Th:CaF$_2$ crystal, by readout of the PMT. The error signal is used as feedback to the EOM to compensate for the long-term drift of the cavity, closing the nuclear clock feedback loop.
\textbf{b}, Frequency comparison to the Yb$^+$ single-ion clock located at BEV is performed by recording the beat signal on a frequency comb, which is stabilized by another high-finesse cavity (not shown) and referenced via a Doppler-compensated fiber link and a second comb to the Yb$^+$ clock laser.
\textbf{c}, Schematic representation of Th:CaF$_2$ with a presumed thorium dimer configuration at the D-center.
\textbf{d}, The Th-229 quadrupole structure level diagram of the D-center, highlighting the investigated clock-transition 5/2 $\rightarrow$ 3/2.}
\label{fig:setup_cw}
\end{figure*}
The development of 148\,nm continuous-wave (CW) laser systems in 2025~\cite{Lal:2025,Xiao:2026}, in combination with highly doped Th-229 CaF$_2$ crystals~\cite{Beeks:2023b,Beeks:2024}, has enabled the measurement of the nuclear transition in absorption spectroscopy~\cite{morawetz2026continuous}. With this, it became possible to probe the Th-229 nuclear transition continuously, without being limited by the long isomeric state lifetime of $\sim$10\,min. At the same time, absorption provides a higher detection efficiency, yielding three orders of magnitude more signal photons per second compared to fluorescence.

An optical clock is based on an atomic transition of narrow linewidth, which is probed by a clock laser with long coherence time. By stabilizing this laser to the atomic resonance, it can inherit the desired stability and reproducibility from the atom. Aiming for millihertz resolution and accuracy, the feedback times in this process typically extend to several seconds, requiring a pre-stabilization of the laser for higher-frequency noise to an optical cavity, which has a suitable short-term stability. With the combination of these two control loops, very stable laser frequencies can be generated and then converted into a time signal through a frequency comb acting as an optical clockwork~\cite{Udem2002}. 

Here, we transfer this concept from an atomic to a nuclear reference transition. While previous studies~\cite{Zhang:2024,ooi2026frequency,morawetz2026continuous} demonstrated that it is possible to excite the Th-229 nucleus using VUV lasers stabilized to external frequency standards, it has not been possible yet to construct a system in which the laser interrogating the Th-229 nuclei is steered by the nuclear transition itself. Therefore, the system presented in this work constitutes the first implementation of a nuclear clock that operates as a stand-alone device. 

\subsection*{Experimental setup}\label{sec2}
The nuclear clock is composed of a cavity-stabilized laser for short-term stability (referred to as the clock laser), and a Th-229 interrogation system providing long-term stability (Fig.~\ref{fig:setup_cw}a). For interrogation of the Th-229 nuclei, a CW VUV laser source is used~\cite{Lal:2025,morawetz2026continuous}. It is comprised of a 1187\,nm frequency-quadrupled commercial laser (Seed Laser, Fig.~\ref{fig:setup_cw}) and a single-pass second harmonic generation (SHG) stage using a randomly quasi phase-matched (RQPM) strontium tetraborate crystal (SBO). The frequency and linewidth of the 1187$\,$nm interrogation laser are determined by an offset phase lock to the clock laser. We modulate the offset lock frequency to perform absorption measurements.
As a clock laser, we use a high-finesse cavity-stabilized external-cavity diode laser (ECDL) operating at 1187\,nm. The stabilization of the clock laser to the cavity uses a Pound-Drever-Hall (PDH) scheme on a sideband generated by an electro-optical modulator (EOM), allowing to shift the frequency of the clock laser independently of the cavity resonances.

The $148\,$nm laser enters a vacuum chamber containing the Th:CaF$_{2}$ crystal and the detector. We use a segment of the X2 crystal~\cite{Beeks:2022}; other pieces of the same ingot were used in~\cite{masuda2019x,Hiraki:2024,Tiedau:2024,Zhang:2024,Higgins:2024,Schaden:2025,ooi2026frequency,hiraki2025laser,guan2026x}. The temperature of the crystal is monitored throughout all measurements; without active stabilization it remains within an interval of $\pm0.5$\,K around 294.7\,K. The detection is based on a photomultiplier tube (PMT) with CsI-photocathode, operating in photon-counting mode, placed behind the crystal for the absorption spectroscopy of the nuclear transitions~\cite{morawetz2026continuous}.

A beat frequency $f_b$ for clock comparison and diagnostics is measured between the 1187\,nm output of the clock laser and the nearest mode of a stabilized infrared frequency comb (Fig.~\ref{fig:setup_cw}b).
The repetition rate of the comb is stabilized by locking the respective nearest comb mode to a high-finesse cavity-stabilized external-cavity laser (ECL) at a wavelength of 1542\,nm. The absolute frequency calibration of the comb is performed with radiation from an ECL at 1542\,nm, provided via a Doppler-compensated fiber link from the Austrian Federal Office of Metrology and Surveying (BEV), where it is referenced to a Yb$^+$ single-ion clock. We use this for comparison measurements between the Yb$^+$ ion clock and the Th-229 nuclear clock. Using the beat signal between the clock laser and the comb, we characterize the clock laser cavity drift and compensate for the dominant linear component of $\sim$200\,mHz/s.

\subsection*{Clock operation}\label{sec3}

In clock operation, the clock laser is stabilized to the interrogated nuclear transition frequency with the help of a feedback loop that corrects for residual instability or drift of the cavity.
To obtain the error signal of the Th-229
absorption line, we modulate the interrogation frequency between two frequencies separated by $f_\text{FWHM}/\sqrt{3}$, with $f_\text{FWHM}$ describing the full width at half maximum (FWHM) of the measured absorption peak, and record the difference between the signals. Scanning is then done by moving the center frequency over the scan range. The resulting curve has a zero crossing at the absorption peak and a near-linear slope around it. For a detailed description and an example of the obtained error signal shape, see the Methods section.

We can neglect absorption saturation effects as we are operating in a regime where the number of Th-229 nuclei in the excited state is much lower than the ground-state population at any time. Therefore, the absorption amplitude $A$ is constant in time. 
The signal-to-noise ratio (SNR) is given by $\text{SNR}\approx \frac{A}{\sqrt{1-A}}\sqrt{\phi}$, where $\phi$ is the number of photons detected per second. With $65\,$pW of VUV laser power reaching the PMT, which has a detection efficiency of $10\,\%$ and $0.75\,\%$ absorption on the Th-229 $5/2\rightarrow3/2$ transition, we achieve $\text{SNR}\approx17$ after $1\,$s of measurement.

In each clock cycle, we perform a measurement at the center frequency determined in the preceding cycle for a given integration time $T$. The result is a point on the previously measured error signal function, which we then invert to get the actual frequency deviation from the zero crossing. This deviation, expressed as infrared frequency, is then added to the current EOM driving frequency in a single adjustment step to shift the laser back to the Th-229 resonance. The actuation of each adjustment step takes about $1\,$s. The size of the step depends on the integration time $T$ and the stability of the high-finesse cavity used in the clock laser.

If $T$ is short, the noise in the detected PMT counts will result in a high uncertainty in the estimated line center. If $T$ is long, the residual drift of the cavity during the measurement also increases this uncertainty. We can choose $T$ such that the SNR of the error signal is high enough to avoid degrading the short-term stability of the clock laser inherited from the cavity.
For $T = 30\,$min, the feedback signal of the nuclear resonance barely impacts the fractional frequency instability at 1$\,$s, which is then comparable to the measured cavity instability of $\sim 10^{-15}$. 
In contrast to fluorescence measurements, absorption spectroscopy also enables us to use feedback times much shorter than the isomer lifetime. With $T = 20\,$s, the short-term stability deteriorates more strongly from the level obtained with the cavity alone, but the laser is locked to the Th-229 frequency on a shorter timescale. This 
enables faster comparison of the Th-229 and Yb$^+$ transition frequencies which extends the accessible mass range for ultralight dark matter searches.
\begin{figure}[t]
\centering
\includegraphics[width=0.5\textwidth]{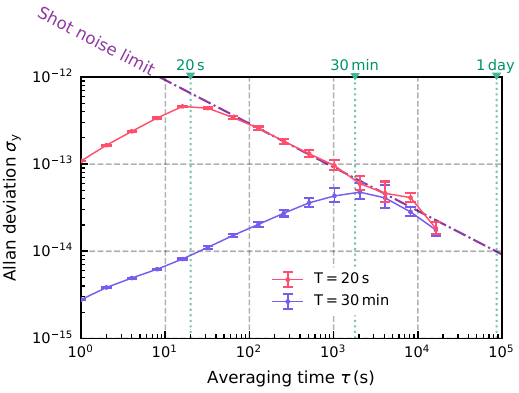}
\caption{\textbf{Fractional frequency instability.}
The fractional frequency instability for different averaging times $\tau$ is shown for two clock operation modes. One mode stabilizes to the $5/2\rightarrow 3/2$ transition with $T=20\,$s between adjustment steps (red curve), the other with $T=30\,$min (blue curve). For $\tau>T$, both curves follow the shot noise trend of the $5/2\rightarrow 3/2$ interrogation.}
\label{fig:Allan}
\end{figure}
\subsection*{Clock performance}

The standard metric for characterizing the stability of a clock is the Allan deviation, the square root of the two-sample variance of consecutive measurements of the normalized clock output frequency.
In Fig.~\ref{fig:Allan} we show 
the overlapping Allan deviation $\sigma_{y}$~\cite{riley2008handbook} plotted versus the averaging time $\tau$. Until averaging times of $\tau\approx T$, it rises for both measurements as $\propto\sqrt{\tau}$. After this, the instability becomes mostly independent of the cavity and is instead determined by the shot noise-limited interrogation of the Th-229 nuclear transition $\sigma_{y}(\tau)\approx (\Gamma /f_0)(1/\text{SNR}) \left(\tau/\text{s} \right)^{-1/2}$. Here, $f_0= 2.0204\cdot10^{15}\,$Hz is the nuclear transition frequency \cite{morawetz2026continuous} and $\Gamma$ is the measured linewidth of $\sim$100\,kHz.

The current state of the power stability of the VUV laser system allows operating the nuclear clock without any intervention for one day. The clock operation shows no effects from the crystal temperature fluctuations $\sim 0.1$\,K, consistent with the temperature dependence of the $5/2\rightarrow 3/2$ transition ($1.8\,$kHz/K at $295\,$K) reported in Ref.~\cite{Higgins:2024}.

\begin{figure*}[ht!]
\centering
\includegraphics[width=1\textwidth]{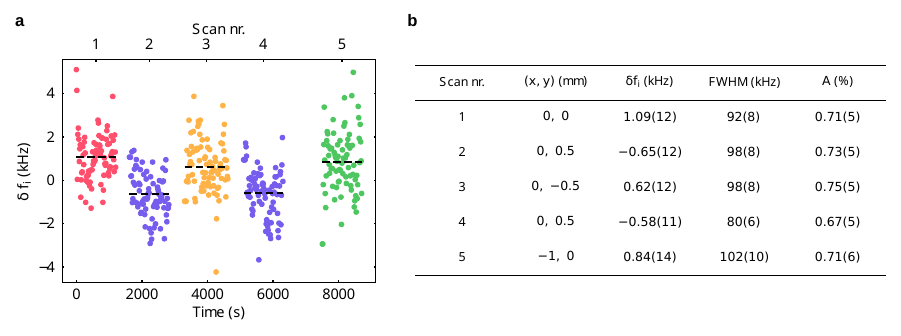}
\caption{\textbf{Crystal position dependent center frequency.} \textbf{a}, Deviations $\delta f_{i}$ of the Th-229 line centers from the frequency average of five consecutive measurements 
at different crystal positions. The second and fourth datasets
(blue) were obtained by operating at the same crystal position. Each point is measured in T = 20\,s.
After t$_{tot}$ = 20\,min of data taking, they average down to the respective dotted black lines. \textbf{b}, The table shows the line center deviation $\delta f_{i}$ for each scanning position. The indicated statistical uncertainties are determined through a Monte Carlo simulation of the clock loop (see Methods). The indicated linewidth at full width half maximum (FWHM) and absorption amplitudes $A$ are extracted from spectroscopy scans performed prior to each clock operation. 
}
\label{fig:reproducibility}
\end{figure*}
While the clock reaches fractional frequency instabilities in the low $10^{-14}$ range within a continuous clock run, we observe that the reproducibility between runs on different days is limited to $\sim 5\cdot10^{-13}$. Before a clock operation run, it is necessary to realign the laser system. In our current setup, this necessarily leads to a small change in the axis and location at which the VUV laser traverses the crystal. Therefore, every clock operation probes slightly different local regions of the X2 crystal.
To test whether this explains the reduced reproducibility, we use an $x-y$ translation stage to perform measurements on five crystal positions. 

We determine the respective line centers through $T=20\,$s measurements over a total duration of $t_\text{tot}=20\,$min on each of the probed crystal positions. As visible in Fig.~\ref{fig:reproducibility}, the deviations $\delta f_{i}$ of the line centers from the frequency average of all five measurements differ by up to $1.7\,$kHz between datasets, which is well above the $0.13\,$kHz spread expected from statistical fluctuations in our measurement, but within the uncertainty reported in previous studies of the line center reproducibility \cite{ooi2026frequency}. Remeasuring at the same crystal position (see scans 2 and 4 in Fig.~\ref{fig:reproducibility}) yields a reproducible line center frequency. We conjecture that local strains induced by structural inhomogeneity causes these frequency shifts. In future implementations of the Th:CaF$_2$-based clock, constraining the optical path through the crystal to a reproducible volume, or improving doping homogeneity and minimizing strain through optimized growth conditions, may enhance reproducibility.

\subsection*{Constraining dark matter couplings}\label{sec4}
\begin{figure*}[ht!]
\centering
\includegraphics[width=\textwidth]{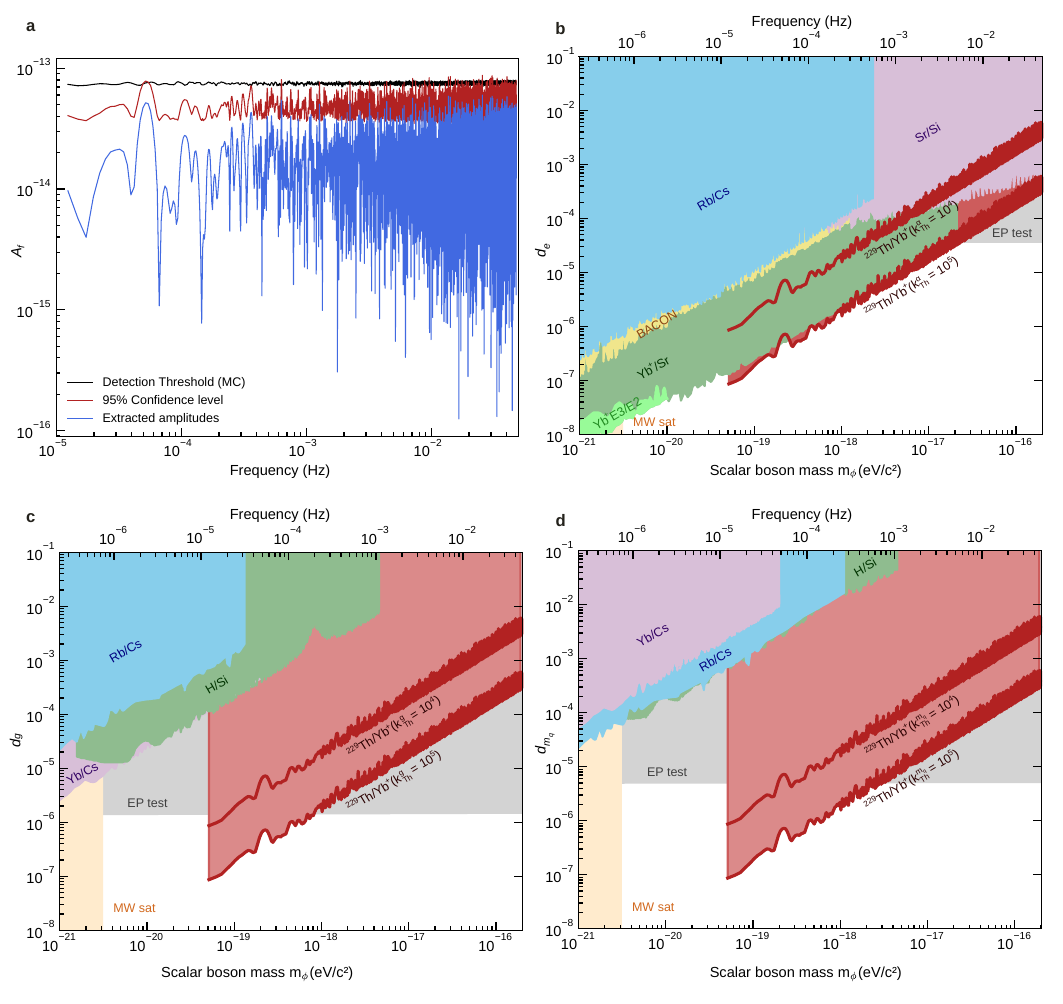}
\caption{\textbf{Dark Matter exclusion plots.} \textbf{a}, Lomb-Scargle periodogram calculated from the $T=20\,$s, $t_\text{tot}\approx23\,$h measurement of the beat frequency of the Th-229 nuclear clock and the Yb$^+$ ion clock. We indicate the detection threshold that was not met at any frequency, as well as the exclusion band. We can exclude with 95\% certainty that no oscillations with amplitudes above this value are present. \textbf{b}, Exclusion plot for scalar bosons with a mass $m_\phi$ and a coupling to photons with a dimensionless coupling constant $d_e$. The shaded areas have been excluded by various clock comparisons \cite{filzinger2023improved, network2020frequency, hees2016searching, kennedy2020precision, kobayashi2022search, banerjee2025oscillating} under the assumption that the scalar boson constitutes the majority of the observed dark matter. Astrophysical observations of satellite galaxies of the Milky Way provide a lower limit on the possible mass of such particles. Equivalence principle tests (EP) \cite{touboul2022microscope, hees2016searching} were able to set limits on new forces mediated by scalar particles. We follow the convention of Ref.~\cite{arakawa2026probing} and give two bands corresponding to sensitivity factors $k_{_\text{Th}}$ of $10^4$ and $10^5$. The sensitivity factors of the Yb$^+$ reference clock of $k_{{\text{Yb}^+}}\approx1$ are negligible in comparison~\cite{Dzuba2003}. \textbf{c}, For the variation of the QCD confinement scale $\Lambda_\text{QCD}$, the sensitivity of the Th-229 transition remains the same while limits from atomic clocks become significantly weaker. Here, the Th-229 nuclear clock is able to test a region of the parameter space that was previously not accessible. \textbf{d}, Also for the variation of the mass of the up and down quarks $m_q$, the Th-229 clock is able to provide new bounds.} 
\label{fig:exclusion plot}
\end{figure*}
To address some of the most pressing open questions in fundamental physics, such as the origin and properties of dark matter, a variety of theories predict the existence of new ultra-light scalar bosons~\cite{jackson2023search}. Even if these bosons only have feeble interactions with the known matter particles, their influence might still be observable in high-precision experiments. For example, couplings to the electromagnetic force, strong force, and quarks would lead to an apparent oscillation of the fine structure constant $\alpha$, the quantum chromodynamics scale parameter $\Lambda_\text{QCD}$, and quark masses $m_q$, respectively, with a frequency corresponding to the mass of the new boson. Because the transition energies in atoms and nuclei depend on these fundamental constants, their values would oscillate too \cite{arvanitaki2015searching}. Here, the Th-229 nuclear transition offers a major advantage: the extremely low energy of the isomer state is a result of a coincidental near-cancellation of MeV-scale electromagnetic and strong nuclear force effects. This makes the transition very sensitive to changes in $\alpha$, $\Lambda_\text{QCD}$, and $m_q$. Due to unknown nuclear parameters, precise values of these sensitivity factors can currently not be calculated \cite{Fine_structure_Beeks_2025,caputo2025sensitivity}, but it is predicted that they will exceed those of the most sensitive atomic clock transitions by multiple orders of magnitude~\cite{Fine_structure_Beeks_2025,caputo2025sensitivity, Flambaum2006,Peik:2021, arakawa2026probing}. Thus, even if the Th-229 nuclear clock does not yet reach the same level of frequency instability as the best atomic clocks, it already yields competitive results in searches for fundamental physics. 

We perform a search for periodic signals in the Th-229 transition frequency on the data from the $T=20\,$s clock operation shown in Fig.~\ref{fig:Allan} (red curve). We can relate the variation in the infrared beat frequency $f_b$ to the variation in the change of the ratio between the Yb$^+$ and Th-229 transition $f_\text{Th}/f_{\text{Yb}^+}$ through $\partial_t\,\text{log}(f_\text{Th}/f_{\text{Yb}^+})=8\,(\partial_tf_b)/f_0$. To search for possible oscillations of this ratio, we compute the best-fit amplitudes $\mathcal{A}_f=\sqrt{4P_f/N_{tot}}/f_0$ using the Lomb-Scargle method~\cite{scargle1982studies}. Here $P_f$ is the power spectrum of the beat frequencies converted into the VUV, and $N_\text{tot}$ is the total number of statistically independent measurements. We are sensitive to oscillations with frequencies between $1/t_\text{tot}$ and $1/T$, where $t_\text{tot}$ is the total measurement time of $\approx 23\,$h. 

To identify the statistical significance of the peaks in the Lomb-Scargle periodogram, we use Monte Carlo simulations of the clock feedback loop to calculate a 5\,\% detection threshold. As no amplitudes exceeded this detection threshold, we establish an upper limit for the amplitude of oscillations in the investigated frequency range (see Fig. \ref{fig:exclusion plot} \textbf{a}, Methods).\\
\indent
The upper limit can further be used to restrict possible couplings to scalar dark matter. The size of the coupling constant $d_e$ for the interaction between a scalar boson $\phi$ and photons relates to the measured power spectrum $P_f$ as (see Methods):\\
\begin{align*}
    d_e&=\frac{1}{k^\alpha_\text{Th}f_0}\cdot\sqrt{\frac{P_f}{N_\text{tot}}}\cdot\frac{ m_\phi}{1.20\times10^{-31}\,\text{eV/c}^2}\\
    &=\frac{\mathcal{A}_f}{k^\alpha_\text{Th}}\cdot\frac{ m_\phi}{2.40\times10^{-31}\,\text{eV/c}^2}\,.
\end{align*}
 The first factor contains the Th-229 specific enhancement through the large sensitivity factor and transition frequency. The second factor indicates how the uncertainty decreases with reduced noise and higher statistics. The final factor leads to the decreasing sensitivities for higher boson masses $m_\phi$ and contains a suppression factor that is the same for any clock measurement. For variations of the fine structure constant $\alpha$, which corresponds to a scalar coupling to photons, the nuclear clock presented in this work imposes limits similar to the best atomic clock comparisons (see Fig. \ref{fig:exclusion plot}). The transition frequencies of atomic clocks are less dependent on nuclear properties, even for those clocks that are based on hyperfine transitions like the Cs clock. Thus, their sensitivities to variations of $\Lambda_\text{QCD}$ and $m_q$ are lower. Here, the nuclear clock presented in this work reaches a factor of 100 to 1000 deeper into the parameter space than earlier experiments with atomic clocks. Compared to Ref.~\cite{arakawa2026probing}, in which measurements of the Th-229 line position and shape over the duration of 10\,months were analyzed, our measurement offers an improvement of around one order of magnitude in the accessible frequency range. The fast feedback of the absorption scheme allows to directly probe masses that could previously only be restricted through line-shape analysis~\cite{fuchs2025searching, arakawa2026probing}. 
In future clock measurements, further enhancements in sensitivity can be expected. The frequency limits of the Lomb-Scargle periodogram are given by $1/t_\text{tot}$ and $1/T$. Therefore, lower frequencies can be reached by operating the clock for longer, whereas higher frequencies are accessible by reducing the cycle time. The sensitivity over the whole frequency range scales with the square root of the number of statistically independent measurements $N_\text{tot}$.\\

Apart from searches for oscillations, we can analyze the frequency ratio $f_\text{Th}/f_{\text{Yb}^+}$ data for possible slow drifts. A linear fit of the $T=20\,$s, $t_\text{tot}\approx 23\,$h data gives us a slope of $(2\pm4)\times 10^{-14}\,$/day, consistent with 0. Using the same sensitivity constants introduced above, this corresponds to a linear drift of $\log\alpha$, $\log\Lambda_\text{QCD}$, and $\log m_q$ of $(2\pm4)\times 10^{-18}[10^{-19}]\,/\text{day.}$

The number in brackets corresponds to the higher sensitivity factor of $10^5$. The aforementioned line-center reproducibility will be a limiting factor when extending this search to the months or years timescale.

\subsection*{Prospects for future solid-state Th-229 nuclear clocks}\label{sec5}

The Th-229 clock performance can be straightforwardly improved by increasing the VUV laser power.
The power limitation of the laser source used in our experiment is defined by a spontaneous domain-inversion structure of the SBO crystal. Only a fraction of the domains contributes to the frequency upconversion of 297\,nm radiation.
One order of magnitude gain in the VUV power can be achieved by using the presented RQPM SBO in an enhancement cavity, which is limited due to optical losses at 297\,nm. An additional enhancement of the VUV power to the hundreds of nW level can be reached by using commercially available 3\,W UV lasers.



Higher VUV power levels can be obtained using four-wave mixing in Cd vapor~\cite{Xiao:2026} where 300\,nW was already reported, or by fabricating patterned SBO~\cite{perlov2024method} and BaMgF$_4$ (BMF)~\cite{Buchter:01} crystals. For a first-order periodically poled nonlinear crystal, a VUV power of $\geq1\,\mu$W can be expected, however, this has not yet been demonstrated due to technical challenges in the sub-micron periodic poling.

The shot noise limited clock instability scales with $\sigma_{y}(\tau)\approx (\Gamma /f_0)(1/\text{SNR}) \left(\tau/\text{s} \right)^{-1/2}$. Beyond increasing the laser power, there are two other directions to improve the clock performance: increasing the absorption and decreasing the linewidth $\Gamma$. 
We operate in a low-saturation regime, where the absorption amplitude $A$ does not change with time. This will be the case as long as the number of photons that are absorbed during the lifetime of the isomeric state is much lower than the number of Th-229 nuclei in the beam path. For the current experiment, around $\sim 3\times10^{8}$ photons are absorbed in 10\,min, while there are $\sim10^{16}$ Th-229 nuclei in the beam path. Therefore, absorption measurements are expected to operate far from saturation even if the available laser power increases significantly with further laser developments.

In Ref.~\cite{ooi2026frequency}, it was found that $\Gamma$ increases with the Th-229 density $\rho_{_\text{Th}}$. The absorption $A$ also depends on $\rho_{_\text{Th}}$ and increases with $A=1-\exp({-\bar{\sigma}L\rho_{_\text{Th}}\chi})$. Here, $L$ is the length of the light path through the crystal, $\chi$ quantifies the percentage of Th atoms that are in a D-center (X2: 58.6\,\%~\cite{hiraki2025laser}), and $\bar{\sigma}$ is the absorption cross section, which itself scales with $\Gamma^{-1}$. These scalings imply that an optimal $\rho_{_\text{Th}}$ exists for which $\sigma_\text{y}$ gets minimized. The shot noise limit resulting from this theoretically optimal doping density and $\chi=1$, is only a factor of 2 lower than the one of our current crystal, X2. Therefore, no major improvements in $\sigma_{_\text{y}}$ can be expected by modifying $\rho_{_\text{Th}}$ in future Th:CaF$_2$ crystals. Through the use of longer crystals and VUV cavities, it should be possible to increase the path length $L$ by a factor of 10, resulting in an order of magnitude higher SNR. With this, and a laser power of $1\,\mu$W, Th-229 doped CaF$_2$ crystals could reach a fractional frequency instability of $\sim10^{-15}/\sqrt{\tau/\text{s}}$.

To go beyond Th:CaF$_2$ one can explore other host crystals: We expect crystals which contain Th as part of the stoichiometric crystal structure, such as ThF$_4$, to have reduced linewidths. In such crystals, Th-229 is not a dopant and thus does not distort the crystal structure, presumably reducing inhomogeneous broadening of the nuclear transition. This is also expected to improve the line center reproducibility.
Another relevant contribution to the linewidth is the magnetic interaction with neighboring F-19 nuclei, causing broadenings in the range of (0.1-1\,kHz)~\cite{Kazakov:2012,Zhang:2024b}. Through the use of spinless solids~\cite{Morgan2025}, even smaller linewidths might become accessible in the future. To benefit from these improvements, the linewidth of future VUV lasers needs to be narrowed down equivalently. Using a conservative linewidth estimate of $\Gamma=1\,$kHz, a crystal with similar doping density and dimensions as the one currently used would yield an optical depth $\bar{\sigma}L\rho_{_\text{Th}}\chi\approx1$. With this, and using a 100\,pW laser with a high stability cavity, a fractional frequency instability of $\sim10^{-16}/\sqrt{\tau/\text{s}}$ could be reached. This would bring the frequency instability of the nuclear clock to the same level as state-of-the-art optical atomic clocks (see, i.e., ~\cite{aeppli2025atomic, filzinger2026multiionopticalclockmathbf5,hb3c-dk28}), while still maintaining the form factor and simplicity of a solid-state clock. Compared to the nuclear clock introduced in this work, this would enable a four orders of magnitude increased sensitivity to the variation of fundamental constants.

\subsection*{Acknowledgments}

We would like to thank Thomas Leder, Martin Menzel, and Andreas Hoppmann for their technical support. We would like to thank Martin Steinel, Burghard Lipphardt, and Nils Huntemann for discussions on the frequency stabilization and optical frequency references. We also thank Dieter Hainz, Monika Veit, Adrian Leitner, Markus Nemetz, and Johannes Sterba from ATI radiation safety for their support in handling radioactive samples. We acknowledge over a decade of support and collaboration with J\"urgen Stuhler, Jan Sch\"afer and the team of Toptica Photonics as well as by Jens Rauschenberger and Rainer H\"orlein from HP Spectroscopy. We thank the National Isotope Development Center of DoE and Oak Ridge National Laboratory for providing the Th-229 used in this work. 

Part of this work has been funded by the European Research Council (ERC) under the European Union’s Horizon 2020 research and innovation programme (Grant Agreement No. 856415) and the Austrian Science Fund (FWF) [Grant DOI: 10.55776/F1004, 10.55776/J4834, 10.55776/ PIN9526523]. We acknowledge support from the \"Osterreichische Nationalstiftung für Forschung, Technologie und Entwicklung (AQUnet project), from the Deutsche Forschungsgemeinschaft (DFG) – SFB 1227 - Project-ID 274200144 (Project B04), and from the Max-Planck-RIKEN-PTB-Center for Time, Constants and Fundamental Symmetries. The project 23FUN03 HIOC [Grant DOI: 10.13039/100019599] has received funding from the European Partnership on Metrology, co-financed from the European Union’s Horizon Europe Research and Innovation Program and by the Participating States. The Vienna team acknowledges funding by Defense Advanced Research Projects Agency (DARPA) under grant number HR0011-25-2-0031. 

\clearpage

\twocolumngrid

\onecolumngrid
\begin{figure*}
\centering
\includegraphics[width=1\textwidth]{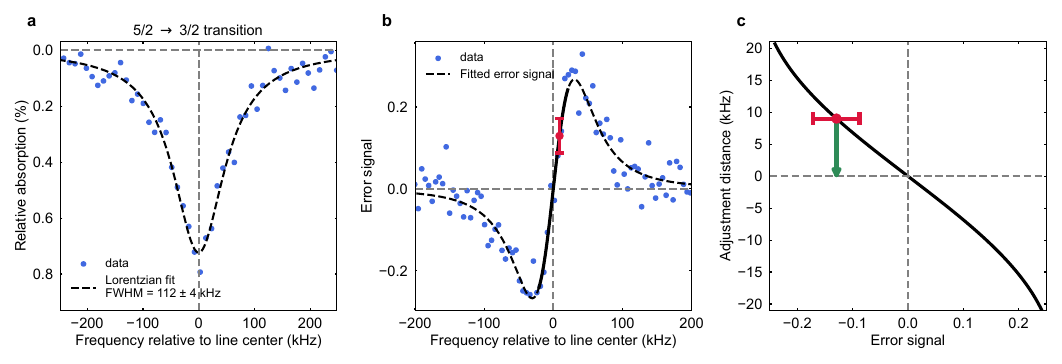}
\caption{\textbf{Calibration and principle of clock operation.} \textbf{a}, Absorption measurement of the $5/2\rightarrow$ 3/2 nuclear clock transition with an integration time $T=3$\,s per data point and an off-resonance frequency-offset of +300\,kHz, expressed as a function of the offset lock frequency translated into the VUV. This is used to determine the linewidth for the error signal scan. \textbf{b}, Error signal measured with an integration time $T=3$\,s per data point and a constant frequency modulation of $\Delta  f_{ol}\,=$\,34\,kHz, from which  the slope is extracted. \textbf{c}, Illustration of a clock frequency adjustment step determined by the fitted error signal shape. }

\label{fig:error_signal}
\end{figure*}

\twocolumngrid

\section*{Methods}\label{meth}
\subsection*{Laser setup}\label{meth:setup}
In the Th-229 interrogation system, the commercial laser (TOPTICA FHG-TA pro) provides an output power of $\approx 500$ mW at a wavelength of 296.8 nm, frequency-quadrupled in two doubling stages and one amplification stage from a diode seed laser at 1187 nm. The final SHG process is based on non-linear frequency conversion in an SBO crystal kept under high-purity N$_2$ atmosphere (purity 5.0). Fundamental radiation still present in the beam after the final SHG step is separated using three dielectric-coated mirrors. For absorption measurements, we use a head-on type PMT (Hamamatsu R6835) mounted inside the vacuum chamber, operated at 2.5 kV. We use the same segment of the X2 sample~\cite{Beeks:2022} as in~\cite{morawetz2026continuous}. The segment in use has a cylindrical geometry with a diameter of 3.1(1) mm and a length of 4.2(1) mm, and is oriented such that the beam traverses along the centerline of the piece. The measured average Th-229 concentration of the segment is 6.6(5)$\times 10^{15}$\,mm$^{-3}$.

The seed laser of the FHG-TA is locked to the clock laser (TOPTICA CLS) via an offset frequency phase lock. A fast photodiode with a bandwidth of 10 GHz measures the beat frequency between the two lasers, which is then mixed with a reference frequency. The fast loop of this lock actuates on the seed laser diode current, and the slow loop is used to steer the grating of the laser diode. By modulating the reference frequency, the laser can be scanned.

Similarly to the actuators of the offset lock, the PDH lock of the clock laser to the cavity is also configured such that the fast feedback loop actuates on the laser diode current, whereas the slow feedback signal of the PDH scheme controls the position of the grating for optical feedback to the laser diode.\\

For clock comparison measurements and drift rate measurements of the cavity, the beat frequency $f_b$ of the clock laser with the frequency comb (Menlo FC1500-250-ULN) is recorded using a dead-time-free frequency counter (K+K FXE) operated in $\Pi$-type counting mode~\cite{rubiola2005, dawkins2007} at a gate time of 1\,s. 
Our method of comparing the thorium clock at TU Wien with the Yb$^+$ clock (TOPTICA TOPTICLOCK)~\cite{Stuhler26} at BEV follows the scheme described in~\cite{morawetz2026continuous}. 
All frequency synthesizers and counters in the Th-229 clock system and the frequency comb are referenced to the 10 MHz signal of a commercial Rb clock (SRS FS725).

To perform an absorption measurement at a specific frequency, we modulate (square wave) the offset lock frequency between the seed laser diode and the clock laser with 10\,Hz between the target frequency and an off-resonance frequency. Each modulation cycle is also reflected in a synchronization signal connected to a time-resolved pulse counter (Swabian Instruments TimeTagger Ultra). The modulation ensures that power fluctuations in the laser output do not affect the measurement. The detection setup consists of a PMT, an RF amplifier, and the pulse counter which bins the arriving pulses from the PMT relative to the last synchronization edge. The absorption is then calculated by subtracting the on-resonance counts from the off-resonance counts and dividing by the off-resonance counts.

\begin{figure*}[t!]
\centering
\includegraphics[width=1\textwidth]{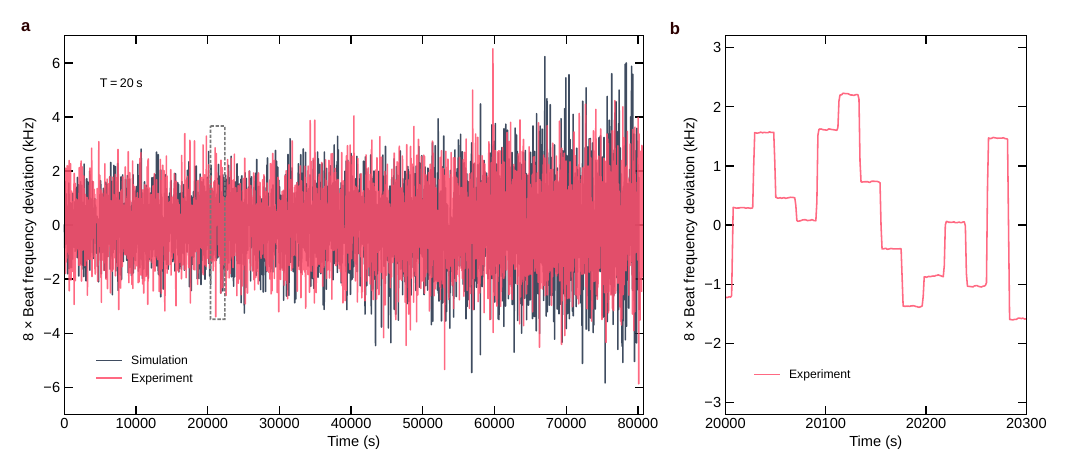}
\caption{\textbf{Beat frequency measurement.} \textbf{a,} The deviation of the IR beat frequency from its mean value for the $T=20\,$s, $5/2\rightarrow3/2$ clock operation mode is shown (red). It is multiplied by a factor of $8$ to indicate its effect on the Th-229 line-center estimate. We additionally show one set of simulated beat frequencies (blue). \textbf{b,} A zoom-in on a shorter timescale ($300\text{ s}$) highlights the signal adjustment steps.}
\label{fig:beats}
\end{figure*}

\subsection*{Clock operation}\label{meth:clockop}

For operating the setup as a clock, we first acquire a single absorption spectrum as shown in Fig.~\ref{fig:error_signal}\textbf{a}. We observe that a Lorentzian line shape fits our absorption measurement data well. We then perform a measurement of the expected error signal (see Fig. \ref{fig:error_signal}\textbf{b}). To measure an error signal, we modulate the offset lock frequency with 10\,Hz between two frequencies $f_\text{low}$ and $f_\text{high}$. The difference $\Delta f = f_\text{high}- f_\text{low }$ is kept constant during the measurement and only the center is swept across the resonance. In our measurements, we used a frequency deviation $\Delta f$ of $ f_ \text{FWHM}/\sqrt{3}$, with $f_ \text{FWHM}$ describing the FWHM of the measured absorption peak (see Fig. \ref{fig:error_signal}\textbf{a}). We can fit this signal with the function: 
\begin{equation*}
    y_\text{fit}(f) =  \frac{A_L}{1 + \left(\frac{ f + \Delta f/2}{\gamma}\right)^2} - \frac{A_L}{1 + \left(\frac{ f - \Delta  f/2}{\gamma}\right)^2}.
\end{equation*}
For the clock operation, the laser frequency is again switched between two frequencies while adjusting the center position with the feedback. For both frequency positions, the counts $c_{\text{low},i}$ and $c_{\text{high},i}$ are recorded, where $i$ is the measurement index for averaging the signal. The applied error signal $E$ can be calculated as 
\begin{equation*}
    E(c) = y^{-1}_\text{fit} \left( \frac{1}{L} \sum_{i = 1}^{L} \frac{c_{\text{high},i} - c_{\text{low},i}}{(c_{\text{high},i} + c_{\text{low},i})/2} \right),
\end{equation*}
with the number of cycles $L$ and the inverse fit function $y^{-1}_\text{fit}(c)$. Fig. \ref{fig:error_signal}c illustrates an adjustment step inferred by the calculation of $E$. With the integration times set in the clock operation, the beat is shifted after every interrogation cycle by the calculated value E, as shown in Fig. \ref{fig:beats}.

\subsection*{Variation of fundamental constants}
The variation of $\alpha(t)$ relates to the variation of the ratio of the Th-229 and ytterbium clock frequencies through: $(k^\alpha_\text{Th}-k^\alpha_\text{Yb})\cdot\partial_t\log\alpha(t)=\partial_t\log(f_\text{Th}/f_\text{Yb})$. Here, $k^\alpha_\text{Th}$ and $k^\alpha_\text{Yb}$ are sensitivity factors that describe how much the respective transition frequency depends on $\alpha$. 
The log-derivative directly corresponds to the measured infrared beat frequency $f_b$ through $\partial_t\log(f_\text{Th}/f_\text{Yb})=8\,(\partial_tf_b)/f_0$, where the factor of $8$ originates from the three frequency doubling steps \cite{Karshenboim2023}.

The anomalously low energy of the Th-229 nuclear transition arises from a coincidental 
near-cancellation of the MeV scale strong force and Coulomb contributions to the binding energies of the involved nuclear states~\cite{Fine_structure_Beeks_2025}.
We therefore expect $k^\alpha_\text{Th}\approx k^{g}_\text{Th}\approx k^{m_q}_\text{Th}$, where $k^{g}_\text{Th}$ and $k^{m_q}_\text{Th}$ are the sensitivity constants that relate between the log-derivatives of the frequency ratio and $\Lambda_\text{QCD}$ and $m_q$ respectively. They by far dominate the sensitivity of the Yb$^+$ reference for which the corresponding sensitivities are $\sim 1$~\cite{Dzuba2003}.
We can determine the amplitude of possible oscillations in the experimental data by computing the Lomb-Scargle power spectrum $P_f$ of the beat signal translated into the VUV. It corresponds to an oscillation amplitude $A_f$ through $A_f = \sqrt{4P_f/ N_\text{tot}} / f_0$, where $N_\text{tot}$ is the total number of datapoints \cite{scargle1982studies}.

Various theories suggest that dark matter could consist of yet undiscovered ultralight scalar bosons~\cite{jackson2023search}. Such bosons would only interact very weakly with other particles and their behavior can be approximately described by a free field: $\phi=\frac{ \sqrt{2(\hbar c)^3\rho_{_\text{DM}}}}{m_\phi c^2}\cos(\frac{m_\phi c^2}{\hbar} t+\delta)$. Here, $\rho_{_\text{DM}}=0.4\,$GeV/cm$^3$ is the observed dark matter density in the Milky Way \cite{jackson2023search}, $m_\phi$ is the unknown mass of the boson, and $\delta$ is some unknown phase. The coupling of such particles to matter can be expressed through the following Lagrangian density~\cite{arvanitaki2015searching,damour2010equivalence,kobayashi2022search}:
\begin{widetext}
\begin{align*} 
    \mathcal{L}\subset -\kappa\phi\Bigg[\frac{d_e}{4\mu_0} F_{\mu\nu}F^{\mu\nu}-\frac{d_g\beta_3}{2g_3}G^a_{\mu\nu}G^{a\mu\nu}+\sum_{q=u,d}\big(d_{m_q}+\gamma_m d_g\big)m_qc^2\bar{\psi}_q\psi_q\Bigg].
\end{align*}
\end{widetext}
Here, we only included the terms that are relevant for this study. $\phi$ is the scalar-, $F_{\mu\nu}$ the $\text{electromagnetic-,}$ $G^a_{\mu\nu}$ the gluon-, and $\psi_q$ the quark-field. $\kappa=\frac{\sqrt{4\pi}}{M_\text{Pl} c^2}$ where $M_\text{Pl}$ is the (unreduced) Planck mass. $\mu_0$ is the vacuum permeability and $\beta_g$ is the QCD beta function that describes the running of the coupling constant $g_3$. $\gamma_q d_q$ describes the anomalous and $m_q$ the bare contribution to the quark mass. The first term leads to a modification of the electromagnetic coupling strength $\alpha\rightarrow \alpha+\alpha \kappa d_e\phi$ \cite{damour2010equivalence}.
The amplitude of this oscillation is given by $\mathcal{A}_f=\frac{\alpha\kappa d_e \sqrt{2(\hbar c)^3\rho_{_\text{DM}}}}{m_\phi c^2}$.
Finally, using the Lomb-Scargle power spectrum $P_f$ of the VUV equivalent of the measured beat signal, we arrive at the following expression for the scalar-photon coupling strength:
\begin{align*}
    d_e=\sqrt{\frac{c^{5}}{\hbar^3}}\cdot\frac{3M_\text{Pl}}{\sqrt{8\pi\rho_{_\text{DM}}}}\cdot\frac{1}{k^\alpha_\text{Th}f_0}\cdot\sqrt{\frac{4P_f}{N_\text{tot}}}\cdot m_\phi.
\end{align*}
Here, we included a factor of 3 to account for possible stochastic fluctuations in the dark matter density \cite{centers2021stochastic}.
The corresponding equations for $\Lambda_\text{QCD}$ and $m_q$ follow completely analogously and only differ by the respective sensitivity factor.

To evaluate the statistical significance of the amplitudes in the Lomb-Scargle power spectrum, we performed Monte Carlo (MC) simulations of the clock feedback loop.
For this, we take $N=T+1$ ($T$ is the integration time in s) random values from the poissonian distribution to simulate the actual experimental cycle. This distribution follows from the SNR of the Lorentzian fit of the measured data at positions $ 
f_\text{est}\pm\frac{\text{FWHM}}{2\sqrt{3}}$. As a next step, we add white noise to these values with an amplitude that matches the noise we observe on the beat frequency. We then average these $N$ values and use the error function to calculate a new estimate for $ f_\text{est}$ on which we repeat the procedure. We further assume a decrease in SNR for each subsequent random value to model the linear decrease in laser power that we experience during the data taking. In Fig.~\ref{fig:beats}\textbf{a}, it can be seen that the simulated beat frequency has the same characteristics as the measured one. We further saw that the power spectrum and Allan deviation of the simulated data fitted well to the experiment.
\\

For the experimental measurement acquired between April 2, 2026 (08:30:15 UTC) and April 3, 2026 (07:04:15 UTC), we extract a single beat frequency value in each cycle to construct a sample of independent frequency ratios. The Lomb-Scargle periodograms of both experimental data and simulations are computed using the Astropy python package \cite{Astropy2022} with psd normalization.

To assess the statistical significance of peaks in the experimental periodogram and account for the look-elsewhere effect, we estimate a 5\% detection threshold. When finding a peak above this threshold, the probability of it being a false detection is less than $p_0 = 5\%$.
Adopting the method of \cite{filzinger2023improved, scargle1982studies}, we perform 1,000 MC simulations of the clock loop and compute the corresponding Lomb-Scargle periodograms. Then, we fit the histogram of the cumulative power at each frequency with an exponential cumulative distribution function of the form: $1 - \exp(-a(P_f - P_0))$. The fit parameters, $a$ and $P_0$, allow the estimation of the detection threshold $P_{f,\text{th}}$ and the corresponding amplitude according to: $P_{f,\text{th}} = P_0 - \frac{1}{a} \ln [ 1 - (1 - p_0)^{1/n_\text{ind}} ]$, where $n_\text{ind} \approx \frac{t_{\text{tot}}}{T}\approx 3800$ is the number of independent frequencies. Here, the estimated threshold fits well to the analytical solution for an exponential distribution, that is expected for white noise. We find no amplitudes exceeding the detection threshold, indicating the absence of DM oscillations in our measurements.

The local 95\% confidence level is determined using 1,000 MC simulations. For each simulation, a random offset is added to the experimental beat frequency at each time step. The offset follows a Gaussian white noise distribution, with a standard deviation $\sigma = 1.34$\,kHz, derived from the experimental beat frequency. The local 95\% confidence level is then defined as the 95th percentile of the amplitude distribution at each frequency, allowing us to exclude, with 95\% certainty, the presence of any oscillations with amplitudes exceeding this limit.

\bibliography{sn-bibliography}
\end{document}